\documentclass[twocolumn,prl,showpacs,nofootinbib]{revtex4}

\usepackage{graphicx}
\usepackage{dcolumn}
\usepackage{bm}
\def\hatn{\bm{\hat n}}

\def\rmB{\mathrm{B}}
\def\rmE{\mathrm{E}}
\def\rmA{\mathrm{A}}

\def\VEV#1{\left\langle #1 \right\rangle}

\newcommand{\beq}{\begin{equation}}
\newcommand{\eeq}{\end{equation}}
\newcommand{\beqa}{\begin{eqnarray}}
\newcommand{\eeqa}{\end{eqnarray}}

\newcommand{\map}{{\mathrm{map}}}
\newcommand{\Npix}{N_{\mathrm{pix}}}

\begin{document}

\title{How to De-Rotate the Cosmic Microwave Background Polarization}

\author{Marc Kamionkowski}
\affiliation{California Institute of Technology, Mail Code 130-33,
Pasadena, CA 91125}

\date{\today}

\begin{abstract}
If the linear polarization of the cosmic microwave background
(CMB) is rotated in a frequency-independent manner as it
propagates from the surface of last scatter, it
may introduce a B-mode polarization.
Here I show that measurement of higher-order TE, EE, EB, and TB
correlations induced by this rotation can be used to reconstruct
the rotation angle as a function of position on the sky.  This
technique can be used to distinguish primordial B modes from
those induced by rotation.  The effects of rotation can be
distinguished geometrically from similar effects due to
cosmic shear.
\end{abstract}


\maketitle

The polarization of the cosmic microwave background (CMB) can be
decomposed into gradient (E mode) and curl (B mode) components
\cite{Kamionkowski:1996ks}.
Primordial density perturbations produce a polarization pattern
that is purely E mode at the surface of last scatter, while
primordial gravitational waves, such as those from inflation,
produce a B mode \cite{Kamionkowski:1996zd}.
There is now an active experimental effort to detect B modes, as
these would constitute a ``smoking gun'' for inflationary
gravitational waves \cite{Bock:2006yf}.

There may, however, be other mechanisms---apart from
gravitational waves---for producing B modes.  The most widely
considered is cosmic shear: the deflection of CMB photons due to
weak gravitational lensing by density perturbations along the
line of sight will convert some of the E mode at the surface of
last scatter to a B mode \cite{Zaldarriaga:1998ar}.  Cosmic
shear of the CMB is no longer the realm of futuristic theorists;
it has recently been detected \cite{Smith:2007rg}.

Another possibility is a rotation of the linear polarization of
the CMB as it travels from the surface of last scatter
\cite{Lue:1998mq}.  This
could occur, for example, if there is a quintessence field that
couples to the pseudoscalar of electromagnetism
\cite{Carroll:1998zi}.  In this case,
the polarization would be rotated by an angle $\alpha$ that is
uniform across the sky.  A fraction $\sin2\alpha$ of the E mode
would thus be converted into a B mode
\cite{Lue:1998mq}.  This B mode could be
distinguished from that due to gravitational waves by the
parity breaking EB and TB cross-correlations that it
produces, as well as by the shape of the TB, EB, and BB power
spectra.  Searches with current data already constrain such a
uniform rotation to be no more than a few degrees
\cite{Feng:2006dp,Komatsu}.

But what if the rotation angle $\alpha$ varies across the sky?
Refs.~\cite{Pospelov:2008gg} have recently proposed models in
which this might occur by virtue of a spatially and time-varying
scalar field coupled to the pseudoscalar of electromagnetism,
and something similar may 
arise from dark-matter magnetic moments \cite{Gardner:2006za}.
(Faraday rotation could also
rotate the polarization \cite{Kosowsky:1996yc}, but this can be
identified with multifrequency maps.)  If the distribution of
rotation angles is symmetric about zero, then there will be a B mode
induced with no parity-breaking TB or EB correlations
\cite{Pospelov:2008gg}.  Can this type of B mode be
distinguished from that due to primordial gravitational waves?

In this {\sl Letter}, I show that a position-dependent rotation of
the polarization induces higher-order correlations in the
temperature-polarization of the CMB.  I then show how these new
correlations can be used to measure the rotation angle
$\alpha(\hatn)$ as a function of position $\hatn$ on the sky.
The observed polarization pattern can then be corrected to
construct the primordial polarization pattern.  It will thus be
possible to distinguish whether a B-mode pattern, if detected,
is primordial or due to a post-recombination rotation.  The
techniques I will discuss can also be used to test an experiment
for systematic artifacts.

The reconstruction algorithm will be similar to that
developed to reconstruct the cosmic-shear field
\cite{Seljak:1998aq} and also to that developed to test for
statistical isotropy (SI) \cite{Pullen:2007tu}.  In fact, the
effects of rotation may be viewed as a possible contaminant for
cosmic-shear maps.  However, the effects of cosmic shear (or SI
violations) and rotation can be distinguished, as (a) rotation
has a different geometric effect on the polarization, and (b)
cosmic shear affects the temperature as well.

The linear polarization at each point $\hatn$ on
the sky is quantified by Stokes parameters $Q(\hatn)$ and
$U(\hatn)$, measured with respect to the $\hat\theta$-$\hat\phi$
axes.  These Stokes parameters are components of a symmetric
trace-free $2\times2$ polarization tensor field ${\cal
P}_{ab}(\hatn)$ which can be expanded in terms of tensor
spherical harmonics as,
\begin{equation}
     {\cal P}_{ab}(\hatn) = \sum_{l=2}^\infty\sum_{m=-l}^l
     \left[ E_{lm}
     Y_{(lm)ab}^{\rmE}(\hatn) + B_{lm} Y_{(lm)ab}^{\rmB}
     (\hatn) \right].
\label{Pexpansion}
\end{equation}
Here, $Y_{(lm)ab}^{\rmE}(\hatn)$ and $Y_{(lm)ab}^{\rm
B}(\hatn)$ are complete sets of basis functions for the gradient
(E mode) and curl (B mode) components of the polarization.
Orthonormality of the basis functions allows us to 
write the expansion coefficients as
\begin{eqnarray}
     E_{lm}&=&\int \, d\hatn {\cal P}_{ab}(\hatn)
                             Y_{(lm)}^{{\rmE} \,ab\, *}(\hatn),
                              \\
     B_{lm}&=&\int d\hatn\, {\cal P}_{ab}(\hatn)
                                      Y_{(lm)}^{{\rmB} \, ab\, *}(\hatn).
\label{eqn:defmoments}
\end{eqnarray}

Suppose now that the polarization pattern at the surface of last
scatter has no B mode and that the polarization at each point
$\hatn$ is rotated by an angle $\alpha(\hatn)$.  In that case,
the observed Stokes parameters will be
\begin{eqnarray}
     \left( \begin{array}{c} Q_{\mathrm{obs}} (\hatn) \\
     U_{\mathrm{obs}} (\hatn) \\ \end{array}
     \right)
     &=& \left( \begin{array}{cc} \cos 2
     \alpha(\hatn) & \sin 2\alpha(\hatn) \\ - \sin 2\alpha(\hatn)
     & \cos 2 \alpha(\hatn) \\ \end{array} \right)
     \left( \begin{array}{c} Q(\hatn) \\ U(\hatn) \\ \end{array}
     \right) \nonumber \\
     & = & 
     \cos2\alpha \left( \begin{array}{c} Q \\ U \\ \end{array}
     \right) 
     +\sin 2\alpha \left( \begin{array}{c} U \\ -Q
     \\ \end{array} \right).
\end{eqnarray}
The concordance of measured TT, TE, and EE power spectra suggest
that $\alpha$ is relatively small, and that the uniform
component of $\alpha$ is small \cite{Feng:2006dp}.  We therefore
work to linear order in $\alpha$.  The change to
the polarization tensor induced by the rotation is then
$\delta {\cal P}_{ab}(\hatn) =2\alpha(\hatn) {\cal P}^{\mathrm
r}_{ab}(\hatn)$, where ${\cal P}^{\mathrm r}_{ab}(\hatn)$ is
rotated from the original polarization ${\cal P}_{ab}$ by
$45^\circ$.  Thus, if ${\cal P}_{ab}$ is pure E mode, then
${\cal P}^{\mathrm r}_{ab}$ is pure B mode and can be written as
\begin{equation}
     {\cal P}^{\mathrm r}_{ab}(\hatn) = 
     \sum_{l=2}^\infty\sum_{m=-l}^l E_{lm} Y_{(lm)ab}^{\rmB}.
\end{equation}

To calculate the curl component induced by rotation, we insert
$\delta {\cal P}_{ab}$ into the expression,
Eq.~(\ref{eqn:defmoments}), for $B_{lm}$.  We then
expand the rotation angle in terms of (scalar) spherical
harmonics $Y_{(lm)}$ as $\alpha(\hatn) = \sum_{LM} \alpha_{LM}
Y_{(LM)}(\hatn)$, to obtain
\begin{eqnarray}
     B_{lm} &=& \sum_{L M}
     2 \sum_{l_2 m_2} \alpha_{L M} E_{l_2 m_2} \int d\hatn
     Y_{(lm)}^{\mathrm{B},ab\,*} Y_{(LM)} Y^{\mathrm
     B}_{(l_2m_2)ab} \nonumber \\
      &=& 2(-1)^m \sum_{LM}\sum_{l_2 m_2}
     \alpha_{LM} E_{l_2m_2} \xi^{LM}_{lm l_2 m_2} H^L_{ll_2},
\label{eqn:Bamplitude}
\end{eqnarray}
where the sum is taken only over $l_2$ values that satisfy
$L+l+l_2=$even,
\begin{equation}
     H^L_{ll'} \equiv \left( \begin{array}{ccc} l & l' & L
     \\ -2 & 2 & 0 \\ \end{array} \right)
     \left( \begin{array}{ccc} l & l' & L
     \\ 0 & 0 & 0 \\ \end{array} \right)^{-1},
\end{equation}
in terms of Wigner-3j symbols, and
\begin{equation} \label{E:xint}
     \xi^{LM}_{lml^\prime m^\prime} =
     \int d\hatn \, Y_{(lm)}^\ast(\hatn)
     Y_{(l^\prime m^\prime)}(\hatn)
     Y_{(LM)}(\hatn).
\end{equation}
We thus see that rotation induces a B mode
\cite{Lue:1998mq,Pospelov:2008gg}.  If
there is a power spectrum for $\alpha_{LM}$, then the CMB power
spectrum $C_l^{\mathrm{BB}}$ can be calculated from
Eq.~(\ref{eqn:Bamplitude}) \cite{Pospelov:2008gg}.

There is also an $O(\alpha)$ change in the E mode induced by
rotation with precisely the same form as
Eq.~(\ref{eqn:Bamplitude}), but with contributions only
from $L+l+l_2$=odd, rather than even.  In the
discussion below, we focus for brevity and clarity on the
induced EB and TB correlations.  However, the entire discussion
applies (with differences to be pointed out below) to induced EE
and TE correlations.

The next step is to consider the correlation of the induced
B mode with the original E mode, as well as with the temperature T.
Recalling that the original E modes have (assuming Gaussian
initial conditions) expectation values $\VEV{E_{lm}^*
E_{l'm'}}=C_l^{\mathrm{EE}} \delta_{ll'} \delta_{mm'}$ (where
$C_l^{\rmE\rmE}$ is the EE power spectrum), the
correlation between the observed E and B modes is
\cite{Scoccola:2004ke},
\begin{eqnarray}
     \VEV{B_{lm} E_{l'm'}^*} & =&  2
     \frac{\alpha_{00}}{\sqrt{4\pi}} C_{l}^{\mathrm{EE}}
     \delta_{ll'} \delta_{mm'} \nonumber \\
     & & + 2 \sum_{L\geq 1}\sum_{M=-L}^{L}
     \alpha_{LM} C_l^{\mathrm{EE}} \xi^{LM}_{lml'm'} H^L_{ll'}.
\label{eqn:EBcorrelation}
\end{eqnarray}
I have split off the $L=0$ term in Eq.~(\ref{eqn:EBcorrelation}) to
show that the result for a uniform rotation angle $\alpha$ is
$\VEV{B_{lm} E_{l'm'}^*}= 2 \alpha C_l^{\mathrm EE}
\delta_{ll'}\delta_{mm'}$, as it should be \cite{Lue:1998mq}.
The sum in Eq.~(\ref{eqn:EBcorrelation}) is taken only over
$L+l+l'$=even.  The expression for the induced correlations
$\VEV{E_{lm} E_{l'm'}^*}$ is the same as that in
Eq.~(\ref{eqn:EBcorrelation}), but with $L+l+l'$=odd, and no
$L=0$ contribution.

Likewise, given that the temperature T and E-mode polarization
are correlated at the surface of last scatter, there will also
be a nonzero TB correlation induced by rotation.  The expression
for $\VEV{T_{lm} B_{l'm'}^*}$ is identical to
Eq.~(\ref{eqn:EBcorrelation}) with the replacement
$C_l^{\mathrm{EE}} \rightarrow C_l^{\mathrm{TE}}$, again for
$L+l+l'$=even.  There will also be TE correlations of the same
form, but with $L+l+l'$=odd and no $L=0$ term.

If $\alpha(\hatn)$ varies with $\hatn$, then $\alpha_{LM} \neq0$
for $L\geq1$, and if so, there will be correlations between
$E_{lm}$ and $B_{l'm'}$ (and $T_{lm}$ and $B_{l'm'}$) of different
$lm$ and $l'm'$.  The existence of these off-diagonal correlations
can be used to measure each of the rotation multipole moments
$\alpha_{lm}$, and thus $\alpha(\hatn)$.
The relevant formalism is similar to that for measuring the
cosmic-shear field \cite{Seljak:1998aq}, or for
searching for SI violations \cite{Pullen:2007tu}, so we can adopt
results from prior work.  To do so, we note that
Eq.~(\ref{eqn:EBcorrelation}) is
identical to Eq.~(A1) in Ref.~\cite{Pullen:2007tu} with the
identification $\mathrm{X}=\mathrm{B}$ (or T in place of B),
$\mathrm{X}'=\mathrm{E}$, and $D^{LM,\mathrm{XX}'}_{ll'} =
\alpha_{LM} C_l^{\mathrm{EE}} H^L_{ll'}$ (or $C_l^{\mathrm{TE}}$
in place of $C_l^{\mathrm{EE}}$ for TB).

Our goal is to obtain the minimum-variance estimator $\widehat
\alpha_{LM}$ that can be obtained from a full-sky polarization
map or a temperature-polarization map.  We suppose that the maps
are provided as a measured temperature $T^\map(\hatn)$ and
Stokes parameters $Q^\map(\hatn)$ and $U^\map(\hatn)$ in $\Npix$
pixels on the sky.  The temperature (polarization) in each pixel
receives contributions from the signal, which is the temperature
(polarization) on the sky smoothed with a Gaussian beam of
full-width half maximum (FWHM) $\theta_{\mathrm{fwhm}}$, and a
Gaussian noise with variance $\sigma_T^2$ ($\sigma_P^2$).  The
power spectra for the map are then $C_l^{\mathrm{A},\map} = |W_l^2|
C_l^{\mathrm{A}} + C_l^{\mathrm{A,n}}$, where
$C_l^{\mathrm{A,n}}$ is the noise power spectrum for
A (e.g., $\mathrm{A}=\{\mathrm{TT,EE,BB,TE,TB,EB}\}$).  These
are $C_l^{\mathrm{TT,n}}=(4\pi/\Npix)\sigma_T^2$,
$C_l^{\mathrm{EE,n}}=C_l^{\mathrm{BB,n}}=(4\pi/\Npix)\sigma_P^2$,
and $C_l^{\mathrm{TE,n}} = C_l^{\mathrm{TB,n}} =
C_l^{\mathrm{EB,n}}=0$. Beam smearing is taken into account with
the window function $W_l=\exp(-l^2\sigma_b^2/2)$, with
$\sigma_b=\theta_{\mathrm{fwhm}}/\sqrt{8\ln 2} =
0.0742\,(\theta_{\mathrm{fwhm}}/1^\circ)$.

We now derive the minimum-variance estimator $\widehat\alpha_{LM}$
that can be obtained from EB correlations; the results for the
estimator that can be obtained from TB correlations will be
identical with the replacement E$\rightarrow$B; TE and EE
estimators are similarly derived.

Following Ref.~\cite{Pullen:2007tu}, the minimum-variance
estimator for each $D^{LM}_{ll'} = 2 \alpha_{LM} C_l^{\mathrm{EE}}
H^L_{ll'}$ that can be obtained from the polarization map is
(see also Refs.~\cite{Hajian:2003qq})
\begin{equation}
     \widehat D^{LM,\map}_{ll'} = (G^L_{ll'})^{-1}\sum_{mm'}
     B_{lm}^\map (E_{l'm'}^{\map})^* \xi^{LM}_{lml'm'},
\end{equation}
with
\begin{equation}
     { 4 \pi (2L+1)} G^L_{ll'} = (2l+1)(2l'+1) \left(
     C^{L0}_{l0l'0} \right)^2,
\end{equation}
where $C^{LM}_{lml'm'}$ is a Clebsch-Gordan coefficient.  

The coefficient $\alpha_{LM}$ can be estimated from measurement
of $D^{LM}_{ll'}$ from each $ll'$ pair through $\alpha_{LM} =
D^{LM}_{ll'}/ \left(2 C_l^{\mathrm{EE}} H^L_{ll'} \right)$.  One
can then average the estimates of $\alpha_{ll'}$ from all of the
$ll'$ pairs.  The trick, though, is to weight these all in a
manner that minimizes the variance to $\alpha_{LM}$.  If each
estimator $\widehat D^{LM,\map}_{ll'}$ were statistically
independent, then we could simply weight by the inverse
variance.  However, things are a bit (though not much) more
complicated.

Each of the estimators $\widehat D^{LM,\map}_{ll'}$ are
statistically independent for different $LM$.  They are also
statistically independent for different $ll'$, {\it except} that
$\widehat D^{LM,\map}_{ll'}$ is correlated with $\widehat
D^{LM,\map}_{l'l}$.  To take this into account, we take $l'\geq
l$ and then consider for $l\neq l'$ EB modes as well as BE
modes.  For $l=l'$, there is only a single variance; for $l'>l$,
there is a $2\times2$ covariance matrix in the EB-BE space.

Write the covariances between the different $\widehat D^{LM,\map}_{ll'}$
as 
\begin{equation}
     {\cal C}^{ll'}_{\rmA\rmA'} \equiv \frac{ G^L_{ll'}}
     {(1+\delta_{ll'})} \VEV{ \widehat D^{LM,\rmA,\map}_{ll'}
     \widehat D^{LM,\rmA',\map}_{ll'}},
\end{equation}
for $\{\rmA,\rmA'\}=\{\rmE\rmB,\rmB\rmE\}$.  For $l=l'$, there is no
distinction between EB and BE; the variance in this case is then
${\cal C}^{ll}_{\rmB\rmE,\rmB\rmE} = \frac{1}{2}\left[
C_l^{\rmB\rmB,\map} C_l^{\rmE\rmE,\map} \right]$.
For $l'>l$, the covariances are ${\cal
C}^{ll'}_{\rmB\rmE,\rmB\rmE} = C_l^{\rmB\rmB,\map}
C_{l'}^{\rmE\rmE,\map}$, ${\cal C}^{ll'}_{\rmE\rmB,\rmE\rmB} = C_{l'}^{\rmB\rmB,\map}
C_{l}^{\rmE\rmE,\map}$, and ${\cal C}^{ll'}_{\rmB\rmE,\rmE\rmB} = C_{l}^{\rmB\rmE,\map}
C_{l'}^{\rmB\rmE,\map}$.  

We now write two estimators, $\widehat \alpha_{LM}^{l=l'}$ and $\widehat
\alpha_{LM}^{l<l'}$, the first coming from EB correlations with
$l=l'$ and the second from those with $l<l'$.  We will then
average them, with inverse-variance weighting, to obtain the
final estimator.  The first is
\begin{equation}
     \widehat \alpha_{LM}^{l=l'} = \frac{\sum_{l} F^L_{ll} (W_l)^2
     \widehat D^{LM,\rm
     map}_{ll} G^L_{ll} / {\cal C}^{ll}_{\mathrm{BE,BE}}}
     {\sum_{l}  \left[F^L_{ll} (W_l)^2
      \right]^2 G^L_{ll} / {\cal C}^{ll}_{\mathrm{BE,BE}}},
\end{equation}
where $F^L_{ll'} \equiv 2 C_l^{\mathrm{EE}} H^L_{ll'}$.  The second is
\begin{equation}
     \widehat \alpha_{LM}^{l<l'} = \frac{\sum_{l'>l} 
     W_l W_{l'} G^L_{ll'}F^{L}_{ll'} \sum_{\rmA\rmA'} \widehat
     D^{LM,\rmA,\map}_{ll'}  \left[ \left({\cal
     C}^{ll} \right)^{-1} \right]_{\rmA\rmA'}}
     {\sum_{l'>l}  \left(W_l
     W_{l'} \right)^2 G^L_{ll'} F^{L}_{ll'}F^{L}_{l'l}
     \sum_{\rmA\rmA'} \left[ \left({\cal
     C}^{ll} \right)^{-1} \right]_{\rmA\rmA'}},
\end{equation}
where the matrix inversion is the in the $2\times2$ EB-BE
basis.  Note that the superscripts A on $\widehat D$ are
necessary in these expressions, as these quantities differ for EB
and BE. The variance to the first estimator is
\begin{equation}
    (\sigma_{\alpha_{LM}^{l=l'}})^{-2} = \sum_l \frac{ \left[
     F^L_{ll} (W_l)^2 \right]^2 G^L_{ll}}{{\cal
     C}^{ll}_{\mathrm{BE,BE}}}.
\end{equation}
The variance $(\sigma_{\alpha_{LM}^{l<l'}})^2$ to the second estimator
is obtained from
\begin{equation} 
     (\sigma_{\alpha_{LM}^{l<l'}})^{-2} =
      \sum_{l'>l} G^L_{ll'} F^{L}_{l'l}
     F^{L}_{ll'} (W_l W_{l'})^2 \sum_{\rmA\rmA'}  \left[ \left( {\cal
     C}^{ll'}\right)^{-1} \right]_{\rmA\rmA'}.
\end{equation}

The final minimum-variance estimator $\widehat \alpha_{LM}$ is
then obtained by averaging, with inverse-variance weighting, the
two estimators above:
\begin{equation}
     \widehat \alpha_{LM} = \frac{\widehat
     \alpha_{LM}^{l=l'}(\sigma_{\alpha_{LM}^{l=l'}})^{-2} +\widehat
     \alpha_{LM}^{l<l'}(\sigma_{\alpha_{LM}^{l<l'}})^{-2}}
     {(\sigma_{\alpha_{LM}^{l=l'}})^{-2}
     +(\sigma_{\alpha_{LM}^{l<l'}})^{-2}}.
\label{eqn:finalestimator}
\end{equation}
The variance $(\sigma_{\alpha_{LM}})^2$ for this estimator is
given by $(\sigma_{\alpha_{LM}})^{-2} =(\sigma_{\alpha_{LM}^{l=l'}})^{-2} +
(\sigma_{\alpha_{LM}^{l<l'}})^{-2}$. 
For small $L$ (e.g., a rotation dipole), the two terms will
contribute comparably to the statistical weight.  For larger
$L$, the $l'>l$ estimator should carry most of the statistical
weight.

Estimators from TB correlations are identical to the EB
estimators discussed above with the replacement
E$\rightarrow$B.  Likewise, there will be EE and TE correlations
similarly induced.  (Things simplify for EE, as the $2\times2$
$\rmA\rmA'$ covariance matrix becomes a single variance.)  The
estimators for $\alpha_{LM}$ for EE and TE and their
variances can be constructed analogously.  There will also be BB
correlations, but they will be higher order in $\alpha$.  There
will be no TT correlations induced, as the rotation does not act
on the temperature.  Since T and E are
correlated in the primordial polarization field, there will be
cross correlations between the estimators $\widehat \alpha_{LM}$
from EB, TB, EE, and TE.  This may be an order-unity effect if the
statistical weights of the various estimators are comparable.
The expressions for the complete covariances are long, and so I
leave them for future work.

Here I have shown that CMB temperature-polarization statistics
can be developed to measure the angle $\alpha(\hatn)$ by which
CMB photons were rotated, en route from the surface of last
scatter, as a function of position $\hatn$ on the sky.  Explicit
formulas to obtain the coefficients $\alpha_{LM}$ in a
spherical-harmonic expansion of $\alpha(\hatn)$ from a full-sky
CMB map were provided.  This technique can then
determine whether B modes, if detected, occur at the
surface of last scatter or are due to a post-recombination
rotation of the polarization.  It is interesting to know that
the rotation angle can be determined from the data, rather than
by assumption.  And if the rotation angle is assumed to be zero,
then the techniques developed here can provide a test for
systematic artifacts in the data. (Ref.~\cite{Hu:2002vu}
suggested tests for systematics along these lines.)

The rotation and cosmic-shear formalisms share some
similarities, and so rotation, if it exists, could show up as an
artifact in a cosmic-shear analysis.  However, the detailed
effects are different and can be distinguished in the data.
First of all, cosmic shear has a different parity than rotation;
a given $L$ mode of the cosmic-shear field correlates 
$l$ and $l'$ modes of E and B, respectively, only if $l+l'+L$ is
odd, while rotation correlates them only for $l+l'+L$ odd.
Furthermore, cosmic shear acts  on temperature and polarization,
while rotation leaves the temperature map unaltered.

The formulas for the estimators $\widehat \alpha_{LM}$
will need to be modified to take into account partial-sky
coverage in a realistic map.  However, it will be
straightforward to adapt the techniques that have been developed
to measure cosmic shear of the CMB on a partial sky to measure
the rotation angle.  Likewise, it is straightforward
to simplify the full-sky analysis performed here to
the flat-sky limit, which may be appropriate for sub-orbital
experiments that map the CMB on a small patch of sky.
Since the formalism to reconstruct the rotation angle resembles
that to determine the cosmic-shear field, there may
be other cosmic-shear techniques that can be adapted for
rotation.  For example, maximum-likelihood techniques
\cite{Hirata:2002jy} may be developed to provide even more
sensitive probes of rotation than the quadratic estimators
discussed here.

Finally, it is of interest to know quantitatively how well the
estimators presented here can be used to construct the rotation angle.
Evaluating the expressions for $\sigma_{\alpha_{LM}}$ for WMAP
values for $\sigma_P$ and $\theta_{\mathrm
fwhm}$, we find for the TB estimator
$\sigma_{\alpha_{LM}}=7^\circ$ for $L=2$ \cite{vera}, consistent
with the current WMAP $1\sigma$ constraint to a uniform
rotation (recalling that $\alpha_{LM}=\sqrt{4\pi} \alpha$ for $L=0$).
This then increases, by about 50\%, to $L=100$.  At higher $L$,
the noise increases due to WMAP's
finite angular resolution.  For WMAP, the values for
$\sigma_{\alpha_{LM}}$ for the EB estimator are seven times
larger, and thus not constraining.  Using values for $\sigma_P$ and
$\theta_{\mathrm fwhm}$ appropriate for the Planck
satellite, we find that the errors to the EB and TB estimators are more
comparable; e.g., $\sigma_{\alpha_{LM}}=1.6^\circ$ and
$1.3^\circ$ for EB and TB, respectively, for $L=2$ rising slowly to
$\sigma_{\alpha_{LM}}=3.4^\circ$ and $1.9^\circ$ for $L=500$.
The noise then
increases rapidly for $l\gtrsim500$, when the correlation angle
becomes smaller than the polarization correlation angle.
More detailed and comprehensive numerical results will be presented
in Ref.~\cite{vera}.

\smallskip
\acknowledgments

I thank A.~Cooray, C.~Hirata, and V.~Gluscevic for
useful comments. This work was supported by DoE
DE-FG03-92-ER40701.


\begin{thebibliography}{}

\bibitem{Kamionkowski:1996ks}
  M.~Kamionkowski, A.~Kosowsky and A.~Stebbins,
  Phys.\ Rev.\  D {\bf 55}, 7368 (1997)
  [arXiv:astro-ph/9611125];
  M.~Zaldarriaga and U.~Seljak,
  Phys.\ Rev.\  D {\bf 55}, 1830 (1997)
  [arXiv:astro-ph/9609170].

\bibitem{Kamionkowski:1996zd}
  M.~Kamionkowski, A.~Kosowsky and A.~Stebbins,
  Phys.\ Rev.\ Lett.\  {\bf 78}, 2058 (1997)
  [arXiv:astro-ph/9609132];
  U.~Seljak and M.~Zaldarriaga,
  Phys.\ Rev.\ Lett.\  {\bf 78}, 2054 (1997)
  [arXiv:astro-ph/9609169].

\bibitem{Bock:2006yf}
  J.~Bock {\it et al.},
  arXiv:astro-ph/0604101.

\bibitem{Zaldarriaga:1998ar}
  M.~Zaldarriaga and U.~Seljak,
  Phys.\ Rev.\  D {\bf 58}, 023003 (1998)
  [arXiv:astro-ph/9803150].

\bibitem{Lue:1998mq}
  A.~Lue, L.~M.~Wang and M.~Kamionkowski,
  Phys.\ Rev.\ Lett.\  {\bf 83}, 1506 (1999)
  [arXiv:astro-ph/9812088];
  N.~F.~Lepora,
  arXiv:gr-qc/9812077.

\bibitem{Carroll:1998zi}
  S.~M.~Carroll,
  Phys.\ Rev.\ Lett.\  {\bf 81}, 3067 (1998)
  [arXiv:astro-ph/9806099].

\bibitem{Feng:2006dp}
  B.~Feng {\it et al.}, 
  Phys.\ Rev.\ Lett.\  {\bf 96}, 221302 (2006)
  [arXiv:astro-ph/0601095];
  T.~Kahniashvili, R.~Durrer and Y.~Maravin,
  Phys.\ Rev.\  D {\bf 78}, 123009 (2008)
  [arXiv:0807.2593 [astro-ph]];
  J.~Q.~Xia {\it et al.},
  Astrophys.\ J.\  {\bf 679}, L61 (2008)
  [arXiv:0803.2350 [astro-ph]];
  J.~Q.~Xia {\it et al.}, 
  Astron.\ Astrophys.\  {\bf 483}, 715 (2008)
  [arXiv:0710.3325 [hep-ph]].

\bibitem{Komatsu}
  E.~Komatsu {\it et al.}  [WMAP Collaboration],
  arXiv:0803.0547 [astro-ph].

\bibitem{Kosowsky:1996yc}
  A.~Kosowsky and A.~Loeb,
  Astrophys.\ J.\  {\bf 469}, 1 (1996)
  [arXiv:astro-ph/9601055].

\bibitem{Pospelov:2008gg}
  M.~Pospelov, A.~Ritz and C.~Skordis,
  arXiv:0808.0673 [astro-ph];
  M.~Li and X.~Zhang,
  arXiv:0810.0403 [astro-ph];

\bibitem{Gardner:2006za}
  S.~Gardner,
  Phys.\ Rev.\ Lett.\  {\bf 100}, 041303 (2008)
  [arXiv:astro-ph/0611684].

\bibitem{Seljak:1998aq}
  U.~Seljak and M.~Zaldarriaga,
  Phys.\ Rev.\ Lett.\  {\bf 82}, 2636 (1999)
  [arXiv:astro-ph/9810092];
  M.~Zaldarriaga and U.~Seljak,
  Phys.\ Rev.\  D {\bf 59}, 123507 (1999)
  [arXiv:astro-ph/9810257];
 W.~Hu,
  Phys.\ Rev.\  D {\bf 62}, 043007 (2000)
  [arXiv:astro-ph/0001303];
  T.~Okamoto and W.~Hu,
  Phys.\ Rev.\  D {\bf 67}, 083002 (2003)
  [arXiv:astro-ph/0301031];
  W.~Hu and T.~Okamoto,
  Astrophys.\ J.\  {\bf 574}, 566 (2002)
  [arXiv:astro-ph/0111606];
  M.~Kesden, A.~Cooray and M.~Kamionkowski,
  Phys.\ Rev.\ Lett.\  {\bf 89}, 011304 (2002)
  [arXiv:astro-ph/0202434];
  L.~Knox and Y.~S.~Song,
  Phys.\ Rev.\ Lett.\  {\bf 89}, 011303 (2002)
  [arXiv:astro-ph/0202286];
  M.~H.~Kesden, A.~Cooray and M.~Kamionkowski,
  Phys.\ Rev.\  D {\bf 67}, 123507 (2003)
  [arXiv:astro-ph/0302536];
  A.~Lewis and A.~Challinor,
  Phys.\ Rept.\  {\bf 429}, 1 (2006)
  [arXiv:astro-ph/0601594];

\bibitem{Pullen:2007tu}
  A.~R.~Pullen and M.~Kamionkowski,
  Phys.\ Rev.\  D {\bf 76}, 103529 (2007)
  [arXiv:0709.1144 [astro-ph]].

\bibitem{Scoccola:2004ke}
  C.~Scoccola, D.~Harari and S.~Mollerach,
  Phys.\ Rev.\  D {\bf 70}, 063003 (2004)
  [arXiv:astro-ph/0405396].

\bibitem{Hajian:2003qq}
  A.~Hajian and T.~Souradeep,
  arXiv:astro-ph/0501001.

\bibitem{Hu:2002vu}
  W.~Hu, M.~M.~Hedman and M.~Zaldarriaga,
  Phys.\ Rev.\  D {\bf 67}, 043004 (2003)
  [arXiv:astro-ph/0210096].

\bibitem{Smith:2007rg}
  K.~M.~Smith, O.~Zahn and O.~Dore,
  Phys.\ Rev.\  D {\bf 76}, 043510 (2007)
  [arXiv:0705.3980 [astro-ph]];
  C.~M.~Hirata {\it et al.}, 
  Phys.\ Rev.\  D {\bf 78}, 043520 (2008)
  [arXiv:0801.0644 [astro-ph]].

\bibitem{Hirata:2002jy}
  C.~M.~Hirata and U.~Seljak,
  Phys.\ Rev.\  D {\bf 67}, 043001 (2003)
  [arXiv:astro-ph/0209489];
  C.~M.~Hirata and U.~Seljak,
  Phys.\ Rev.\  D {\bf 68}, 083002 (2003)
  [arXiv:astro-ph/0306354].

\bibitem{vera}
  V.~Gluscevic, A.~Cooray, and M.~Kamionkowski, in preparation.

\end{thebibliography}
\end{document}